\documentclass[prl,twocolumn]{revtex4}
\usepackage{bm}
\usepackage{graphicx}
\usepackage{amsmath}
\usepackage{eufrak}
\usepackage{color}
\usepackage{upgreek}
\usepackage{subfigure}
\usepackage[unicode=true,colorlinks=true,citecolor=blue]{hyperref}
\usepackage{placeins}
\usepackage{lipsum}

\newcommand{\nix}[1]{}

\begin{document}

\title{Giant terahertz photoconductance of tunneling point contacts}

\author{
M. Otteneder,$^1$ Z. D. Kvon,$^{2,3}$ O. A. Tkachenko,$^{2}$ 
V. A. Tkachenko,$^{2-4}$ \\ A. S. Jaroshevich,$^2$ E. E. Rodyakina,$^{2,3}$ 
 	A. V. Latyshev,$^{2,3}$ and S. D. Ganichev$^1$
}

\affiliation{$^1$Terahertz Center, University of Regensburg, 93040 Regensburg, Germany}

\affiliation{$^2$A.V. Rzhanov Institute of
Semiconductor Physics, Novosibirsk 630090, Russia}

\affiliation{$^3$Novosibirsk State University,
Novosibirsk 630090, Russia} 

\affiliation{$^4$Novosibirsk State Technical University, Novosibirsk
630073, Russia}

\begin{abstract}
We report on the observation of the giant photoconductance 
of a quantum point contact (QPC) in tunneling regime excited 
by terahertz radiation. Studied QPCs are formed in a GaAs/AlGaAs 
heterostructure with a high-electron-mobility two-dimensional electron gas. 
We demonstrate that irradiation of strongly negatively biased QPCs by 
laser radiation with frequency $f = 0.69$~THz and intensity 50~mW/cm$^2$
results in two orders of magnitude enhancement of the QPC conductance.
The effect has a superlinear intensity dependence and 
increases with the  dark conductivity decrease. It is also 
characterized by strong polarization and frequency dependencies.
We demonstrate that all experimental findings can be well 
explained by the photon-mediated tunneling through the QPC.
Corresponding calculations are in a good agreement with the experiment. 
\end{abstract}

\maketitle

\section{Introduction}

The quantum point contact (QPC), discovered in 1988~\cite{Wees88,Wharam88}, is one of the most remarkable quantum devices in condensed matter physics~\cite{Buttiker90,Houten96}. QPCs offer an elegant way to investigate one-dimensional phenomena in electronic transport by the electrostatic squeezing of a two-dimensional electron gas (2DEG) and become attractive for fundamental research of charge transport in mesoscopic conductors and numerous applications. Due to the fact that characteristic energies of QPCs are of the order of meV and can be electrically tuned by gate voltages, they become an important candidate for frequency-sensitive terahertz detection~\cite{Dorozhkin}. A conversion of high frequency (terahertz/microwave) electric fields in a \textit{dc} electric current has been demonstrated in QPCs operating in various regimes and attributed to either electron gas heating~\cite{Song10,Wyss93,Karadi94} or rectification due to the nonlinearity of the QPC current-voltage characteristics~\cite{Song08,Janssen94,Karadi94}. Qing Hu considered a feasibility of the photon-assisted quantum tunneling transport~\cite{Hu93} but follow up experiments did not provide an evidence for this effect because the observed photoresponse has been dominated by the electron gas heating effects~\cite{ch2Wyss93p1522,Wyss95}. 
We note that almost all works on the terahertz/microwave response of QPCs were devoted to either the open regime with $G > 2e^2/h$, in which the conductance quantization in units of $2e^2/h$ is detected, 
or the pinch-off regime at $G \leq 2e^2/h$. 

First studies of the photocurrent in QPCs operating in tunneling regime ($G \ll
2e^2/h$) have been carried out most recently and demonstrated quite unexpected results - a giant microwave photoconductance in response to 100~GHz radiation~\cite{Levin15}. 
The experiments reveal 
an enhancement of the QPC conductance by more than two orders of magnitude for a rather small microwave power density of about 10~mW/cm$^2$. 
The effect has been detected for traditional split-gate QPCs as well as for QPCs with a specially designed bridged-gate. 
In the latter structures, a photoresponse about ten times larger than in the ordinary split-gate QPC has been detected. The photoconductivity and the difference between the two types of gates have been shown to be caused by the influence of microwaves on the steady-state electron distribution function near the tunnel contact, i.e. a specific form of electron gas heating~\cite{Levin15}.

Here, we demonstrate that the giant photoconductance in QPCs operating in tunneling regime can also be obtained by applying radiation of 
substantially higher frequencies in the terahertz (THz) range.  
We show that at THz frequencies, the effect is caused by the photon-assisted tunneling and exhibits a strong polarization and frequency dependency. 
We present a microscopic theory of the photon-mediated tunneling in QPCs,
which is in good quantitative agreement with the experimental findings.

\section{Samples and methods}
\subsection{Samples}

\begin{figure}[t]
\includegraphics[width=\linewidth]{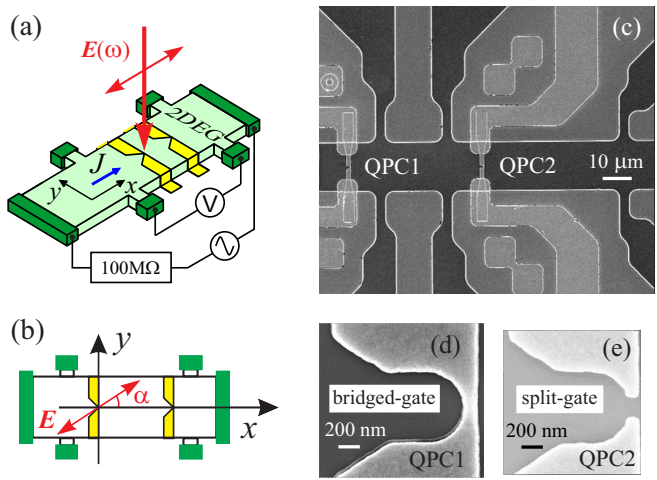}
\caption{ (a) and (b) Sketch of Hall bar samples with 
two QPC structures obtained by gates 
placed on two parts of the sample.
The double arrow indicates the electric field 
vector ${\bm E}(\omega)$ of normally incident THz radiation.
The orientation of ${\bm E}(\omega)$ with respect to the $x$-axis
is described by the azimuthal angle $\alpha$, see panel (b).
(c) Microphotograph of the gated part of the sample.
(d) and (e) Zoomed images of the bridged-gate QPC 
and a traditional split-gate QPC, respectively.
 }
\label{fig_1}
\end{figure}

 Our samples were fabricated on the basis of modulation 
doped GaAs/AlGaAs heterostructures with a 2DEG. Several samples have 
been prepared from two different wafers. 
The first wafer, 
referred to as \#A
in this paper, has a carrier density and 
mobility of the 2D electrons at liquid helium temperature
of about $n_s = 7 \div 8 \times 10^{11}$~cm$^{-2}$ 
and $\mu = 1.5 \times 10^6$~cm$^2$/Vs, respectively. 
This mobility corresponds to a mean free path of 30~$\upmu$m, 
which substantially exceeds the QPC size being of the order of 100~nm. 
The second wafer (\#B)
has the density $n_s = 5 \div 6\times 10^{11}$~cm$^{-2}$ and a much lower 
mobility $\mu = 4 \times 10^5$~cm$^2$/Vs 
(the corresponding mean free path is about 4~$\upmu$m). 
Figures~\ref{fig_1}(a) and (b)
show schematically quantum point contacts with a traditional split-gate and 
bridged-gate placed on two parts of a Hall bar sample. Microphotographs 
of the gated parts of the sample are shown in Figs.~\ref{fig_1}(c)-(e).
The bridged-gate QPC consists of a single piece of 
metal with a semi-elliptical narrowing, see Fig~\ref{fig_1}(d).

\begin{figure}[t]
\includegraphics[width=\linewidth]{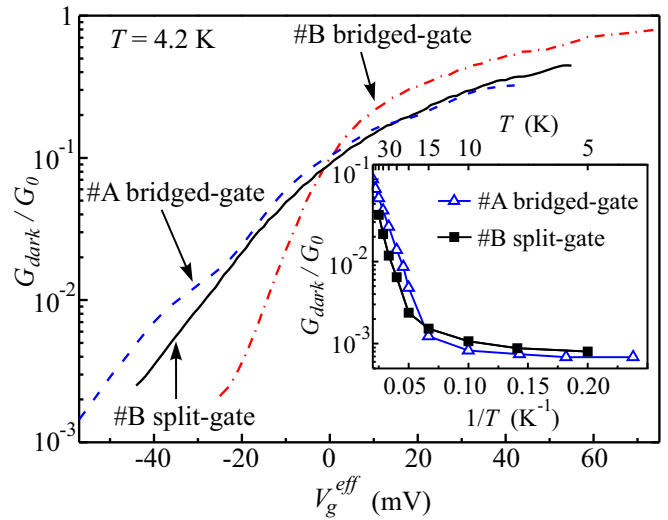}
\caption{  Dependence of the relative dark conductance 
$G_{dark}(V_g^{eff})/G_0$ on the effective gate 
voltage $V_g^{eff} = V_g - V_g{(0.1 G_0)}$.
Here, the values of $V_g$ corresponding to $G = 0.1 G_0$ are 
$-2.23$~V (bridged-gate sample \#A),
$-2.05$~V (split-gate sample \#B) and 
$-1.80$~V (bridged-gate sample \#B).
%
The dependencies sre measured at $T=4.2$~K. 
The inset shows the temperature dependencies 
obtained 
at $V_g^{eff}=  -78$~mV 
for  the bridged-gate sample~\#A
and $V_g^{eff} = -61$~mV  
for the split-gate sample~\#B.
}
\label{fig_2}
\end{figure}

The gates are fabricated on 
the surface of the heterostructure using electron beam lithography with 
a distance of about 90~nm between the gate and the 2DEG. 
The resistance $R = 1/G$ was measured 
using the electric circuit shown in Fig.~\ref{fig_1}(a)
and conventional lock-in technique with 
a frequency of $3$~Hz and currents $J = (10^{-10} \div 10^{-8})$~A.
Figure~\ref{fig_2} shows  the gate voltage dependence of the normalized 
dark conductance $G_{dark}(V_g^{eff})/G_0$ obtained for $J = 10^{-10}$~A. 
Here, $G_0= 2e^2/h$ is the conductance quantum. 
The data obtained for the non-illuminated \#A and \#B structures 
reveal that at a temperature of $T = 4.2$~K and in the used 
range of gate voltages, the conductance $G(V_g)$ 
is much smaller than $G_0$, 
i.e. all QPCs operate in the tunneling regime. 
Note that while the overall characteristics of the QPCs remain the same, 
values of the gate voltage corresponding to the conductance $G_0$
depend on the cooldown procedure and differ from sample to sample.
This is ascribed to cooldown dependent charge trapping in the insulator. 
To compare the measurements taken at different sample
cooldowns we plot the data as a
function of the effective gate voltage $V_g^{eff} = V_g - V_g({0.1 G_0})$ with $V_g({0.1 G_0})$ being the gate voltage at which the conductance is equal to $0.1G_0$.

The regime of complete tunneling of a QPC, with the resistance up to several 
M$\Omega$ is realized for samples
%
cooled below $T \approx 10$~K. 
This can be seen in the temperature dependence 
of the conductance shown for the bridged-gate QPC 
(made of wafer \#A) and the split-gate one (made of wafer \#B) in the inset in 
Fig.~\ref{fig_2}. 
It is seen that the temperature dependencies are similar for both kinds of samples, demonstrating exponential growth for $T > 15$~K (with activation temperature of about 60~K) and saturation of the conductance at low temperatures. The latter gives an evidence for the tunneling regime in the latter case. 
For the similar value of the saturated conductance $G_{dark} = 7\times 10^{-4} G_0$ we estimate the barrier height being of about 5 and 4.5~meV for bridged-gate sample~\#A and split-gate sample~\#B, respectively. 
Note that in spite of the different parameters of the 2DEG and different QPC shapes, we obtain almost the same barrier heights
for the same dark conductance.

\subsection{Methods}

To measure the terahertz photoconductance of the QPCs we used a THz gas 
laser~\cite{Kvon2008,DMSPRL09} optically pumped by a CO$_2$ laser~\cite{jiangPRB2010}. Radiation with frequency 
$f = 0.69$ and 1.63~THz (wavelengths of $\lambda = 432$ and 184~$\upmu$m) have been 
obtained using CH$_2$O$_2$ and CH$_2$F$_2$ gases, respectively. 
The corresponding photon energies $\hbar\omega$ are 2.85 and 6.74~meV, respectively. 
All experiments are performed at normal incidence of radiation, see Fig.~\ref{fig_1}(a), and
a temperature of $T = 4.2$~K. The normal incidence is used to exclude 
other possible photoresponses caused by photogalvanic or photon drag effects, 
which for this geometry are forbidden by symmetry 
arguments~\cite{3authors,Belkov283p2003,ivchenko05a}.
The radiation was focused onto a spot of about 2~mm diameter. 
The spatial beam distribution has an almost Gaussian profile, measured by a 
pyroelectric camera~\cite{Ganichev1999,Ziemann2000}. The radiation intensity $I$ on 
the QPC structures was about 50~mW/cm$^2$ and 
200~mW/cm$^2$ for $f = 0.69$ and 1.63~THz, respectively. The polarization 
plane of the radiation has been rotated by an azimuthal angle 
$\alpha$ using a lambda-half crystal quartz plate~\cite{removalSiGe}. 
The photoresponse, similar to the transport experiments, 
is measured by applying the electric circuit shown in Fig.~\ref{fig_1}(a) 
for currents $J = (10^{-9} \div 10^{-8})$~A.

\begin{figure}[t]
\includegraphics[width=\linewidth]{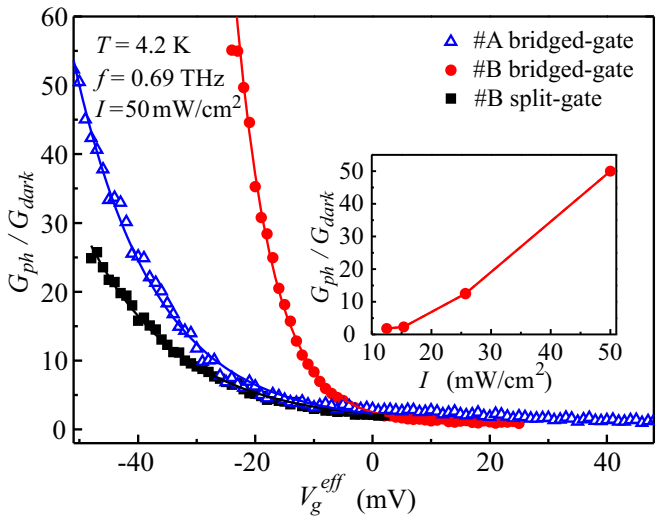}
\caption{ Dependencies of the relative terahertz photoconductivity $G_{ph}/G_{dark}$ 
on the effective gate voltage $V_g^{eff}$. The data are obtained for wafers \#A and
\#B
with different gate configurations by
applying radiation with $f = 0.69$~THz 
and intensity $I = 50$~mW/cm$^2$. 
Solid curves show fits according to $G_{ph}/G_{dark} = A \exp(- V_g^{eff}/a)$, where $A$ and $a$
are the fit parameters.
The inset displays the superlinear dependence of the photoconductivity on radiation intensity $I$.
}
\label{fig_3}
\end{figure}

\section{results}

First, we discuss the results obtained at a radiation frequency $f = 0.69$~THz
with the electric field vector oriented along $x$-direction, i.e. normally to the gate stripes. 
Irradiating the structure with the THz beam, we detected a photoconductive signal corresponding to the conductivity increase, which drastically raises for large negative gate voltages, see Fig.~\ref{fig_3}. 
The latter
corresponds to the dark conductance decrease, see Fig.~\ref{fig_2}. 
The figure reveals that at the lowest dark conductance the photoconductance 
exceeds the dark one by almost two orders of magnitude. Figure~\ref{fig_3} also shows that the 
photoconductive response  $G_{ph}/G_{dark}$  is substantially larger for lower 
mobility samples, see data in Fig.~\ref{fig_2} 
for bridged-gated samples \#A and \#B. Furthermore, its magnitude is 
larger for the bridged-gate QPC than for the split-gate structure. 
This difference has been previously observed in the microwave range on 
similar structures~\cite{Levin15}.
Studying the photoresponse as a function of the radiation power, we observed that it is characterized by a superlinear dependence,
see the inset in Fig.~\ref{fig_3}. 

\begin{figure}[t]
\includegraphics[width=\linewidth]{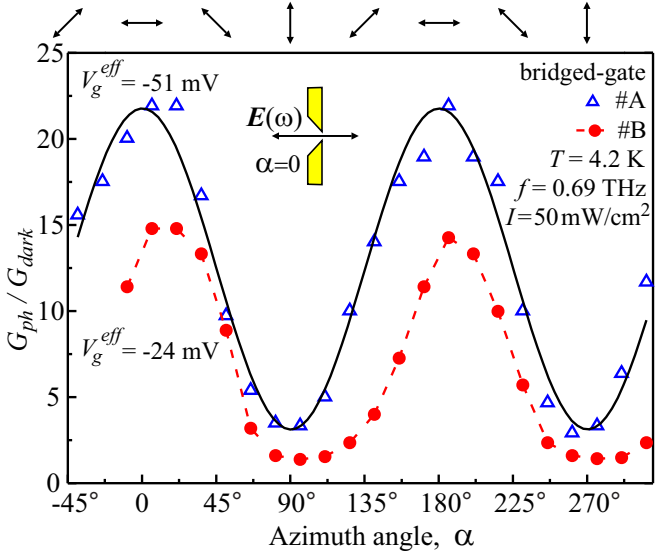}
\caption{ Dependencies of the terahertz photoconductance $G_{ph}/G_{0}$ on the orientation of the terahertz radiation electric field vector $\bm E$ 
for bridged-gate samples \#A
(blue triangles) and 
\#B
(red circles). 
The data are obtained for the effective gate voltages $V_g^{eff}=-51$~mV (sample \#A)
and -24~mV (sample \#B) corresponding to the values of $G_{dark}/G_0 = 0.0018$ and 0.003, respectively.  
The solid line is a fit according to $G_{ph}/G_{dark} \propto (\cos \alpha)^2$.
The inset shows the orientation of the vector $\bm E$ corresponding to the signal maximum.
Arrows on top illustrate the orientations of the radiation electric field vector for several values of $\alpha$.
}
\label{fig_4}
\end{figure}

Exploiting the advantage of THz laser radiation, which,
in contrast to radiation in the microwave frequency range, 
permits to carry out accurate polarization experiment~\cite{Herrmann16},
we investigated the variation of the photoresponse as a function of the 
orientation of the radiation electric field vector. 
Figure~\ref{fig_4} shows the dependence of the photoconductance $G_{ph}/G_{dark}$
on the azimuthal angle $\alpha$ measured for bridged-gate samples \#A and \#B.
The figure reveals that the photoresponse can be well fitted by $G_{ph}/G_{dark} \propto \cos^2 \alpha$ and achieves its maximum for a radiation electric field vector oriented perpendicular to the gate stripes (${\bm E} \parallel x$).

\begin{figure}[t]
\includegraphics[width=\linewidth]{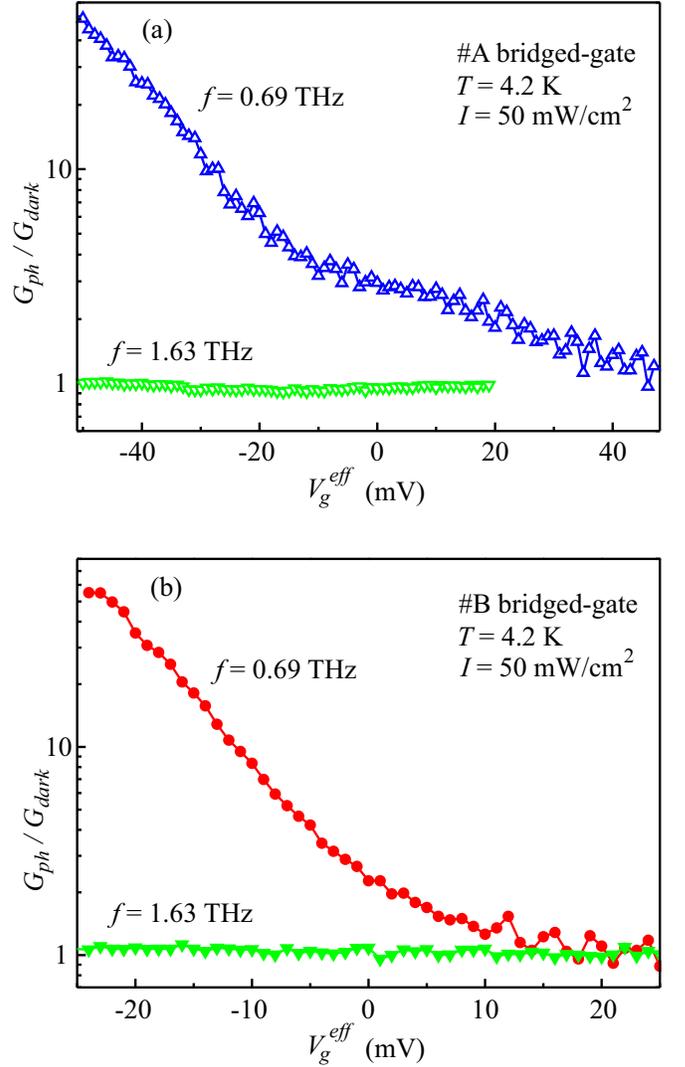}
\caption{  (a) and (b) Dependencies of the photoconductance $G_{ph}/G_{dark}$
on the effective gate voltage $V_g^{eff}$ measured for two radiation frequencies
in bridged-gate samples \#A and \#B, respectively.
}
\label{fig_5}
\end{figure}

Now, we turn to the photoresponse obtained for higher radiation frequency.
Increasing the frequency by about 2.5 times, we observed a drastic (by more than one order of magnitude) reduction  
of the photoconductance $G_{ph}/G_{0}$. This is shown in Fig.~\ref{fig_5}
for bridged-gate samples \#A and \#B excited by radiation with 
frequencies $f=0.69$ and 1.63~THz. Furthermore, for higher frequencies, the photoresponse becomes almost independent on the effective gate voltage.

\section{Discussion}
\label{discussion}

The giant terahertz photoconductance of a tunneling point contact
and its behavior upon variation of radiation polarization, frequency and intensity 
can be well described by the model 
of coherent photon assisted tunneling in QPCs developed in Ref.~\cite{Tkachenko16}.

\begin{figure}[t]
\includegraphics[width=\linewidth]{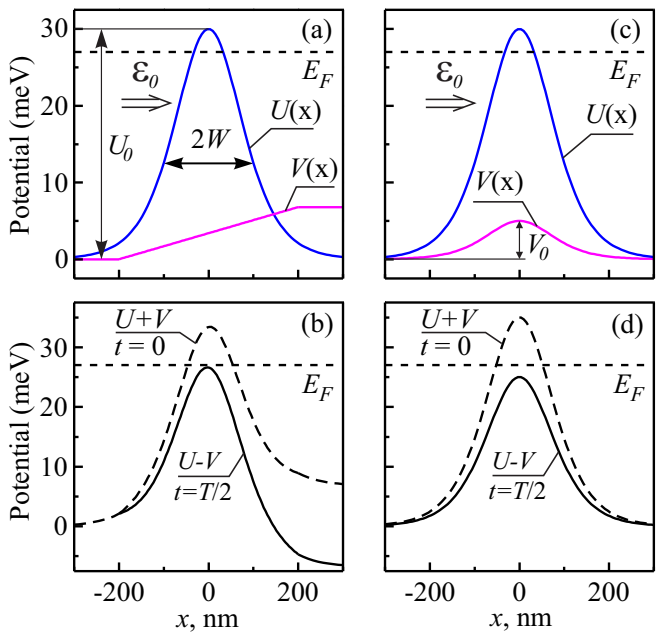}
\caption{ Schematic drawings of the QPC potential barriers  $U(x)$
in the plane of the QPC and their modifications
upon action of the radiation electric field given by 
$U(x) + V(x)\cos(\omega t)$. 
(a) Electrostatic potential barrier in 
$x$-direction (solid blue line) and radiation electric force 
$e E_{0,x}(\omega) \cos(\omega t) = [dV/dx] \cos(\omega t)$ acting on the electrons for 
one half of a period at $|x|<2W$ (solid magenta line). 
(b) Modification of the potential barrier due to 
the $E_x(\omega)$-component of the radiation field is
shown for two instants in time, $t=0$
and $t=T/2$, shifted by half a period.  
(c) Electrostatic potential barrier in $x$-direction and dynamic
barrier with the height $V_0$ caused by $E_z(\omega)$-component of the radiation electric field. 
(d) Change of the potential $U(x)+V(x)\cos(\omega t)$ due to the $E_z(\omega)$-component shown for 
two instants in time, $t=0$ and $t=T/2$, shifted by half a period.
}
\label{fig_7}
\end{figure}

Let's first consider, at a qualitative level, the enhancement of the electric current in 
a QPC operating in the tunneling regime under the action of normally incident linearly polarized
microwave/terahertz radiation. 
The effect of the radiation field is twofold: 
(i) It causes an additional force in the direction of the tunneling current 
       due to the $x$-component of the terahertz electric field ${\bm E}(\omega)$ and 
(ii) it reduces the tunneling barrier 
        due to the $E_z(\omega)$-component of the electric field originating from the near field of 
diffraction\,\cite{r9,GanichevPrettl}. 
The model of the potential modification and the corresponding enhancement of tunneling is shown in Fig.~\ref{fig_7}. 
It is assumed that the QPC in the tunneling regime can be considered as a 
one-dimensional barrier $U(x)=U_0/{\rm ch}^2(x/W)$ (blue solid curve), where $U_0$ is 
the barrier height and $W$ is its characteristic width.  
Figure~\ref{fig_7}(a) and (b) illustrate the influence of the $E_x$-field component resulting in a 
force $e E_{0,x}(\omega) \cos(\omega t) = [dV(x)/dx]\cos(\omega t)$ 
applied along $x$-direction shown in Fig.~\ref{fig_7}(a)
by the magenta line for one half of a period of the wave. The force results in a 
time-dependent potential $U(x) + V(x)\cos(\omega t)$, which increases the tunneling current for one half of a period of the radiation field and decreases it for the other half.
The corresponding potentials are shown in Fig.~\ref{fig_7}(b) by solid and dashed curves. Obviously the effect is maximal for the radiation polarized along 
the current, i.e. in $x$-direction, as detected in our experiment.
The second mechanism, considering the action of the potential 
$V(x) = V_0 / {\rm ch}^2(x/W)$, is illustrated in Fig.~\ref{fig_7}(c) and (d). It is based on the 
reduction of the tunneling barrier $U(x)$ due to the $z$-component of the electric field. 
In the vicinity of the QPC formed by the spiked split-gate and for a radiation electric field 
oriented along $x$-direction, i.e. normal to the gate stripes, near field diffraction
results  in a field $E_z$ directed along $z$-direction for one half of a period
and $-z$ for the other. 
The corresponding time-dependent variation of 
the  potential
is shown in Fig.~\ref{fig_7}(d) by solid  and dashed curves.
The reduction of the potential barrier yields the increase of the 
tunneling current and, consequently, the conductance. For an electric field oriented along $y$-direction, 
the $E_z$-component becomes more complex and has opposite signs for opposite sides of the 
spiked gates forming the QPC and the effect vanishes.

Figure~\ref{fig_7} and the above discussion present a semiclassical  
description of the radiation induced tunneling in QPCs, which illustrates the basic physics of the phenomena. 
This model, however, is valid only for rather low frequencies 
at which the characteristic time of electron tunneling through the potential barrier $U(x)$ is smaller than $1/\omega$~\cite{Buttiker82}.
For arbitrary frequencies the effect can be  
described by the theory of the  photon-assisted transmission~\cite{Coon85} 
adopted for QPC in the tunneling regime in Ref.~\cite{Tkachenko16}.
In the framework of this model, the Schr\"odinger equation with 
the time-dependent potential $U(x) + V(x) \cos(\omega t)$ is solved numerically, where $U(x)$ and $V(x)$ are 
approximated by piecewise constant functions 
(step approximation) as Coon and Liu suggested in Ref.~\cite{Coon85}.
For incident electrons having energy ${\cal E}_0$, the coefficient $D$ of total multichannel 
transmission was determined by taking into account the essential channels 
${\cal E} ={\cal E}_0 \pm n \hbar \omega$ with absorption/emission of $n$ photons. 
The conductance was found by the Landauer formula at zero temperature: 
$G = G_0 D({\cal E}_0)$. As experiment shows, at a fixed $V_g$ 
the dark conductance does not depend on temperature in the region $T < 10$~K (see the inset to Fig.~\ref{fig_2}). Therefore, we performed all the calculations 
for $T=0$.

\begin{figure}[t]
\includegraphics[width=\linewidth]{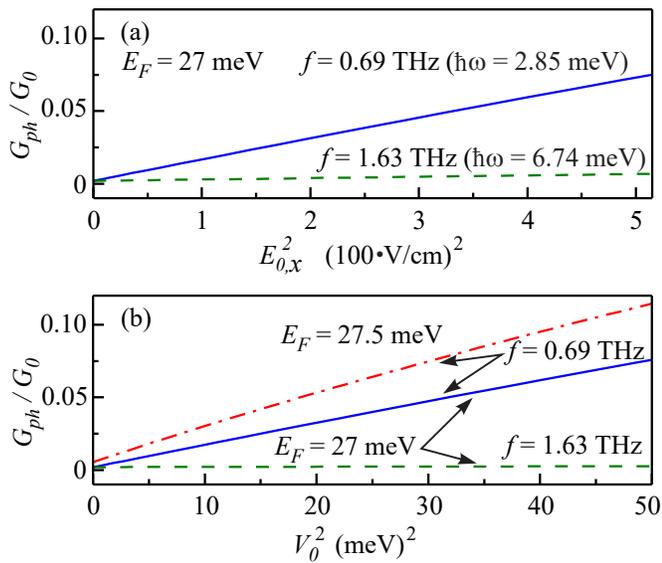}
\caption{ Calculated power dependencies of the photoconductance 
in a tunneling QPC in response to terahertz radiation of two frequencies.
Panels (a) and (b) show dependencies on the magnitude of the radiation 
electric field strength $E_{0,x}^2$ and dynamic
barrier amplitude  $V_0^2$, respectively. 
Calculations are carried out using $T = 0$~K, $E_F$= 27~meV, $U_0 = 30$~meV, and $W = 85$~nm.
Panel (b) shows also the result of calculations for $f=0.69$~THz and a slightly 
larger Fermi energy $E_F$= 27.5~meV (red dotted-dashed line).
}
\label{fig_8}
\end{figure}

\begin{figure}[t]
\includegraphics[width=\linewidth]{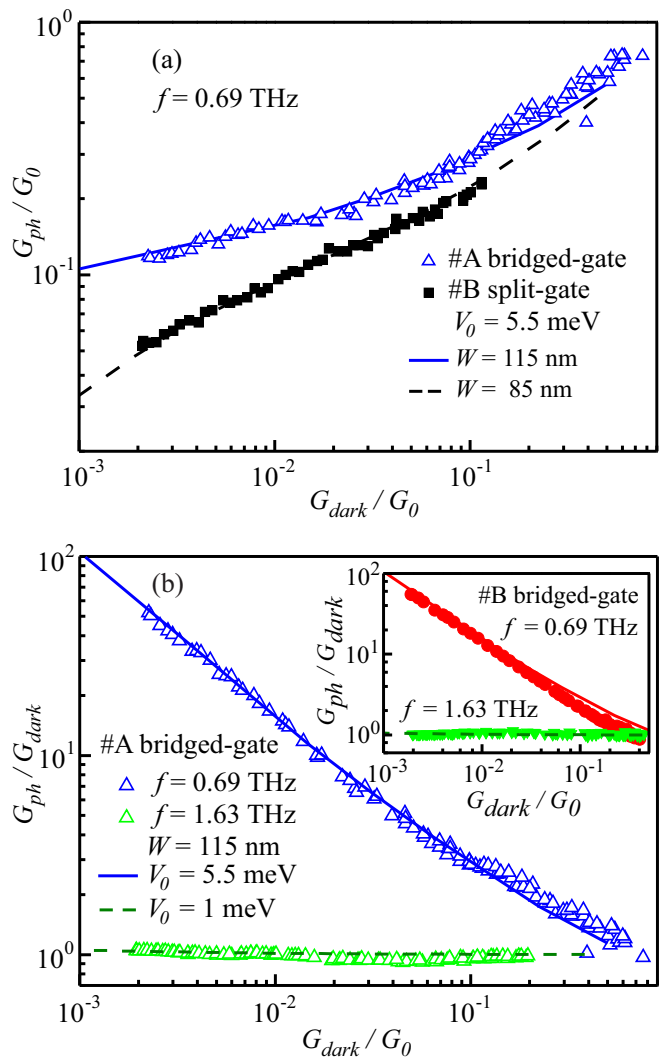}
\caption{
Theoretical and experimental dependencies of
$G_{ph}/G_{0}$ (a)  and 
$G_{ph}/G_{dark}$ (b) on the 
normalized dark conductivity $G_{dark}/G_{0}$.
The data in the panel (a) are shown for
bridged-gate sample \#A and split-gate sample \#B excited by radiation with
$f = 0.69$~THz and intensity $I = 50$~mW/cm$^2$. Calculated curves are obtained with 
$U_0 = 30$~meV, $V_0=5.5$~meV,
$W = 115$~nm (sample \#A) and 85~nm (sample \#B). 
The data in panel (b) are presented for bridged-gate sample ~\#A excited by
radiation frequencies $f=0.69$ and 1.63~THz (intensity $I = 50$~mW/cm$^2$). Curves are 
calculated using $U_0 = 30$~meV, $W = 115$~nm 
and potentials $V_0=5.5$~meV ($f=0.69$~THz) and 1~meV ($f=1.63$~THz). The inset shows 
dependencies for the bridged-gate sample ~\#B.
}
\label{fig_9}
\end{figure}

Figure~\ref{fig_8} presents the results of calculations 
carried out in the framework of the above model. It shows the dependence of the conductance $G/G_0$ on the 
square of the radiation electric field $E_{0,x}^2$, panel (a), and the corresponding potential $V_0^2$, panel
(b), calculated for frequencies 0.69 and 1.63~THz. 
Here, the ratio of $E_{0,x}/V_0$ has been scaled 
in the way that the magnitudes of the conductance for $f=0.69$~THz 
at the highest $V_0$ matches that at the highest $E_{0,x}$-field.
Comparison of these two panels demonstrates that both mechanisms yield
almost the same behavior of $G(E_{0,x}^2)$, panel (a), and  $G(V_0^2)$, panel (b). 
The calculated dependence describes the experimental one well as shown in the inset in Fig.~\ref{fig_9}(b)
for the bridged-gate sample \#B.
A good agreement is also obtained for the dependence of the 
normalized photoconductance $G_{ph}/G_0$ and $G_{ph}/G_{dark}$ 
on the dark conductance $G_{dark}/G_0$, see Fig.~\ref{fig_9}. 
We note that plotting $G_{ph}$ against 
$G_{dark}$ allows to compare experiment and 
modeling independently of the parameters of the 2DEG, the kind of structure, or the shape of gate. 
In particular, the calculations reveal a drastic suppression of the photoresponse both for the transition of the QPC from tunneling to open regime and for the change of frequency in the tunneling regime from 0.69 to 1.63~THz.
This is in full agreement with
the experimental results, see Figs.~\ref{fig_5} and \ref{fig_9}(b).
The higher photoconductance for 0.69~THz 
($\hbar \omega =2.85$~meV) is explained by transitions to the channels 
${\cal E} = {\cal E}_0 \pm n \hbar\omega$, $n=0, 1$. These channels belong to the tunneling mode 
and thus have classical turning points with a high probability density of 
discovering the electron, whereas for 1.63~THz ($\hbar \omega =6.74$~meV) 
the channel ${\cal E}_0 + \hbar \omega$ already belongs to the open mode with larger 
delocalization, smaller probability density in the barrier, and, therefore, 
reduced transition efficiency.

Finally, we note that the fitting of the model to the experiment is 
mainly determined by two parameters --- the characteristic width $W$ 
of the QPC barrier and the intensity of irradiation described by 
$V_0$ (or $E_{0,x}$). The values for optimal fitting $W = 115$
and 85~nm for bridged-gate and split-gate samples 
are close to the lithographic size of the gates. 
The amplitude of the high frequency potential $V_0 = 5.5$~meV corresponds 
to the extreme value $dV(x)/edx=423$~V/cm or $E_{0,x} = 176$~V/cm, which is almost 70 or 30~times larger than the maximum value 
calculated for the incoming plane wave ($E_0=6.14$~V/cm 
at the radiation intensity $I=50$~mW/cm$^2$). 
This observation agrees with previous studies relying on the near field of diffraction
for which a strong enhancement of the field amplitude in the vicinity of a metal edge 
(in our case the spiked gates forming the QPC) has been reported~\cite{GanichevPrettl,ratchet_graphene,PRB2017}.

\section{Possible role of the electron gas heating}

Finally, we discuss a possible role of electron gas heating in the observed phenomena. 
First, we address the effect of the electron temperature increase. A substantial contribution of this mechanism can be ruled out based on the polarization dependence
of the photoconductivity and on estimations of the electron temperature provided by independent measurements on the macroscopic part of the sample.
A strong variation of the photoresponse with rotation of the radiation electric field vector $\bm E$, see Fig.~\ref{fig_4}, indicates that the tunneling current across the QPC has a maximum for $\bm E$ oriented parallel to the current  ($\alpha = 0$) and vanishes for $\alpha = 90^\circ$.
For the electron gas heating mechanism, however, the increase of the tunneling probability is caused by the rise of the 2DEG temperature due to Drude-like radiation absorption
and any polarization dependence is not expected. 
The conclusion that an increase of the electron temperature does not play a substantial role in the photoconductive response is also supported by the estimation of  the effect of electron gas heating. For that we performed measurements on 
a macroscopic part of the sample which does not contain the QPC structure. Comparing the temperature dependence of the conductivity with and without continuous-wave
 THz radiation (not shown),
we observed that at the highest power level used in this work the 2DEG temperature increases 
only by 1~K or even less. This increase of the electron temperature is not sufficient to explain the observed increase by one to two orders of magnitude, see Fig.~\ref{fig_3}.

 While the increase of the electron temperature seems to be unlikely to be responsible for the observed signal, the mechanism based on the influence of terahertz radiation 
on the steady-state electron distribution function near the tunnel contact considered in~\cite{Levin15} is not excluded by the above arguments. This mechanism should be enhanced for the radiation electric field vector oriented along $x$-direction, as observed in the experiment, and does not depend on the electron temperature. At the same time, the magnitude of the photoconductivity in this case is also proportional to the Drude absorption, which is characterized by a weaker frequency dependence as that detected in our experiments, see Fig.~\ref{fig_5}. According to this model, changing the frequency from 0.69 to 1.63~THz should result in a decrease of the photoresponse magnitude by about 5 times. In experiments, however, this factor is about 50, i.e. by one order of magnitude larger. Thus, the theory of Ref.~\cite{Levin15} yielding a good agreement for microwave frequencies does not explain a strong frequency dependence of the terahertz radiation induced photoconductivity in the QPC structure.

\section{Summary}

To summarize, we have observed a giant terahertz photoconductive response of the QPC operating in the tunneling regime. Experimental observations are in good agreement with the theory of the photon-mediated 
tunneling. The observed effect  enlarges the family of photon/phonon assisted tunneling
phenomena, previously detected in semiconductor systems for  superlattices~\cite{GanichevPrettl,ch2Guimaraes93p3792,ch2Keay95p4098}, resonance tunneling diodes~\cite{ch2Chitta92p432,ch2Drexler95p2816},
quantum cascade laser structures~\cite{FaistQCL}, quantum
dot systems~\cite{ch2Kouwenhoven94p3443} and semiconductors doped with deep impurities~\cite{ch2Ganichev93p3882,ch2Ganichev98p2409}.
The observed change in conductivity by more than two orders of magnitude in response to
rather week terahertz radiation with a power of the order of several milliwatts demonstrate that a QPC in tunneling regime can be considered a good candidate for detection of terahertz radiation. 
As for the future work, the most challenging task
is the search for the step-like dependency of the photoconductive response, known for the photon-assisted tunneling in 
superconductors~\cite{ch2Dayem62p246,ch2Tien63p1647,shapiro63} and semiconductor superlattices~\cite{GanichevPrettl,ch2Guimaraes93p3792,ch2Keay95p4098}. The theory developed in Ref.~\cite{Tkachenko16} 
reveals that this behavior is expected for an even deeper tunneling regime and lower temperature.

\section*{Acknowledgments}
\begin{acknowledgments}

SDG thanks S.~Mikhailov and V.~Popov for fruitful discussions. 
The financial support of this work by German Science Foundation DFG (SFB~1277-A04), RFBI (grants N 17-02-00384 and N 16-38-00851),   and the Volkswagen Stiftung Program is acknowledged.
\end{acknowledgments}


\end{document}